\documentstyle[12pt]{article}

\newcommand{\ee}{{\mathbf{e}}}

\begin{document}

\title{Klein-Gordon equations for energy-momentum of relativistic
particle in rapidity space}
\author{
Robert M. Yamaleev\\
Joint Institute for Nuclear Research, LIT,Dubna, Russia\\
Universidad Nacional Autonoma de M\'exico, M\'exico.\\
Email:yamaleev@jinr.ru } \maketitle

\begin{abstract}

The notion of four- rapidity is defined as a four-vector with one time-like  and three space-like coordinates. It is
proved, the energy and momentum defined in the space of four-rapidity obey  Klein-Gordon equations constrained by the
classical trajectory of a relativistic particle. It is shown, for small values of a proper mass influence of the
constraint is weakened and the classical motion gains features of a wave motion.

\end{abstract}

\section{Introduction}

The goal of this paper is to prove that energy and momentum
defined in the space of four- rapidity obey the differential
equation similar the Klein-Gordon equation. This method is based
on a {\it key-formula} connecting the fraction with an exponential
function the argument of which is proportional the difference
between numerator and denominator. The energy and momentum can be
defined either as functions of the hyperbolic angle, or as
functions of the circular angle. In a covariant formulation we
arrive to concept of rapidity expressed as a four-vector the
time-like part of which is presented by the hyperbolic angle. The
circular angle is extended into three quantities functionally
depending of the hyperbolic angle.
 It is shown, the energy and
momentum defined in a such way satisfy the Klein-Gordon equations
written in four-dimensional space with Minkowskii signature.

\section{Key-formulae linking an exponential function with ratio of two quantities}

{\bf 2.1 Parametrization of relativistic evolution with respect to
hyperbolic angle}.

The  dynamical variables of the relativistic particle, the energy
$p_0$, the momentum $p$ and the proper mass satisfy a mass-shell
equation \cite{Barut}
$$
p_0^2=p^2+m^2,  \eqno(2.0)
$$
where the speed of light $c$ taken in unit $c=1$.

Our construction is based on the Key-formula which establishes
some natural interrelation between the ratio of a pair of
quantities and an exponential function.

Consider general form of the complex number given by unit $\ee$ obeying the quadratic equation \cite{Yamaleev4}:
$$
\ee^2-2p_0\ee+p^2=0, \eqno(2.1)
$$
with distinct positive real roots  $x_1,x_2$, so that,
$$
2p_0=x_1+x_2,~~p^2=x_1x_2.  \eqno(2.2)
$$
The coefficients $p_0,p^2$ are real numbers and $p_0^2> p^2$. The
solutions of equation (2.1) are defined by
$$
x_1=p_0+m,~~x_2=p_0-m,~~m=+\sqrt{p_0^2-p^2}. \eqno(2.3)
$$
The normal form of the matrix obeying equation (2.1) is given by
$$
E=\left( \begin{array}{cc}
0&-p^2\\
1&2p_0
\end{array} \right). \eqno(2.4)
$$
Consider an evolution generated by matrix $E$. Write the Euler
formula
$$
\exp(E\phi)=E~g_1(\phi)+I~g_0(\phi), \eqno(2.5)
$$
$I$-is a unit matrix. Diagonal form of this matrix equation
consists of two equations
$$
\exp(x_2\phi)=x_2~g_1(\phi)+g_0(\phi),~~
\exp(x_1\phi)=x_1~g_1(\phi)+g_0(\phi). \eqno(2.6)
$$
Notice, the modified cosine-sine functions $g_0(\phi),~g_1(\phi)$
depend also of coefficients $p_0,p^2$. Form the following ratio
$$
\exp((x_2-x_1)\phi)=\frac{x_2~g_1(\phi)+g_0(\phi)}{x_1~g_1(\phi)+g_0(\phi)}=\frac{x_2+D}{x_1+D},
\eqno(2.7)
$$
where
$$
D=\frac{g_0(\phi)}{g_1(\phi)}. \eqno(2.8)
$$
Let $\phi=\phi_0$ be the point where $g_0(\phi_0)=0$. Then,
$$
\exp((x_2-x_1)\phi_0)=\frac{x_2}{x_1}. \eqno(2.9)
$$
From (2.3) it follows
$$
m=\frac{1}{2}(x_2-x_1),~~p_0=\frac{1}{2}(x_1+x_2),~p^2=x_1x_2.
\eqno(2.10)
$$
Let $\phi=\phi_0$ be an initial point. Then according with (2.9)
we conclude that the roots and the coefficients of equation (2.1)
$x_1,x_2$ and $p_0,p$ are functions of $\phi_0$, however the
difference $2m=x_2-x_1$ does not depend of $\phi_0$:
$$
x_2(\phi_0)=\exp(m\phi_0)\frac{m}{\sinh(m\phi_0)},~
x_1(\phi_0)=\exp(-m\phi_0)\frac{m}{\sinh(m\phi_0)}, \eqno(2.11)
$$
Use these formulae in (2.9). Then,
$$
\exp((x_2-x_1)\phi_0)=\frac{p_0+m}{p_0-m}. \eqno(2.12)
$$
Consequently, we have the following dependence $p_0,p$ of
$\phi_0$:
$$
p_0(\phi_0)=m\coth(m\phi_0),~~p(\phi_0)=\frac{m}{\sinh(m\phi_0)}.
\eqno(2.13)
$$
In Refs.\cite{Yamaleev1,Yamaleev2,Yamaleev3} formula  (2.9) has
been denominated as {\it Key-formula}.

{\bf 2.1 Parametrization of evolution with respect to periodic
angle}.

Now, consider general complex algebra with generator $\ee$ obeying
the quadratic equation
$$
\ee^2-2p\ee+p_0^2=0. \eqno(2.14)
$$
which differs from (2.1) by transposition of the coefficients
$p_0$ and $p$. Since $p_0^2>p^2$, two solutions of equation (2.14)
are given by complex conjugated numbers:
$$
y_2=p+im,~~y_1=p-im,~~m=+\sqrt{p_0^2-p^2}. \eqno(2.15)
$$
Exponential function at solutions of this equation is defined by expansions
$$
\exp(y_2\theta)=y_2~f_1(\theta)+f_0(\theta),~~\exp(y_1\theta)=y_1~f_1(\theta)+f_0(\theta),
\eqno(2.16)
$$
where functions $f_0(\theta),f_1(\theta)$ depend of coefficients
$p,p_0^2$.  Form the following ratio
$$
\exp(i2m\theta)=\frac{y_2~f_1(\theta)+f_0(\theta)}{y_1~f_1(\theta)+f_0(\theta)}= \frac{y_2+F}{y_1+F}
$$
$$
= \frac{p+im+F}{p-im+F}, \eqno(2.17)
$$
where
$$
F=\frac{f_0}{f_1}.  \eqno(2.18)
$$
Let $\theta=\theta_0$ be the initial point where
$f_0(m\theta_0)=0$. Then, formula (2.17) is reduced into the
following relationship
$$
\exp(i2m\theta_0)=\frac{p(\theta_0)+im}{p(\theta_0)-im}.
\eqno(2.19)
$$
From this formula it follows
$$
 p(\theta_0)=m\cot(m\theta_0),~~ p_0(\theta_0)=m\frac{1}{\sin(m\theta_0)}.   \eqno(2.20)
$$
The roots $y_1,y_2$ also are functions of $\theta_0$,
$$
y_1=\exp(-im\theta_0)\frac{m}{sin(m\theta_0)},~
y_2=\exp(im\theta_0)\frac{m}{sin(m\theta_0)}. \eqno(2.21)
$$

 Thus, we obtained two representations for the energy- momentum. The first one is done via hyperbolic
trigonometric functions,
$$
p_0(\phi)=m\coth(m\phi),~~~p(\phi)=m\frac{1}{\sinh(m\phi)},
\eqno(2.22)
$$
and the other one is defined by ordinary periodic trigonometric
functions
$$
p_0(\theta)=m\frac{1}{\sin(m\theta)},~~~p(\chi)=m\cot(m\theta).
\eqno(2.23)
$$
In the both representations the arguments of the trigonometric
functions are proportional to mass $m$.

Since formulae (2.22) and (2.23) are related to same physical
quantities, we come to the next relationships between hyperbolic
and periodic trigonometric functions
$$
\tanh(m\phi)=\sin(m\theta),~\mbox{or},~\sinh(m\phi)=\tan(m\theta).
\eqno(2.24)
$$
Notice, when $m=0$, $\phi=\theta$.

The relationships between $\phi$ and $\theta$  can be presented also as follows
$$
\exp(m\phi)=\frac{1+\sin(m\theta)}{1-\sin(m\theta)}=\frac{1+\tan\frac{m\theta}{2}}{1-\tan\frac{m\theta}{2}},\eqno(2.25a)
$$
$$
\exp(im\theta)=\frac{1+i\sinh(m\phi)}{1-i\sinh(m\phi)}=\frac{1+i\tanh\frac{m\phi}{2}}{1-i\tanh\frac{m\phi}{2}}.
\eqno(2.25b)
$$

Also, it is important to notice that the differential relationship
between variables $\theta$ and $\phi$ coincides with the
definition of the velocity:
$$
\frac{d\theta}{d\phi}=\frac{dr}{dt}=\frac{p}{p_0}.  \eqno(2.27)
$$

These formulae express a general interrelation between periodic
and hyperbolic trigonometry. Let $\triangle ABC$ be a right-angled
triangle with right angle at $C$. Denote the sides by $a=BC,b=AC$,
the hypotenuse $AB$ by $c$. If we make a geometrical motion by
moving the point $A$ along line $AC$, then this motion changes the
sides $c,b$, but remains invariant the side $a$. The angle $A$ can
be used in quality of parameter this evolution. In accordance with
the {\it Key-formula} (2.19) we write
$$
\frac{b+ia}{b-ia}=\exp(2ia\theta),~~~b=a\cot(a\theta),~~c=a\frac{1}{\sin(a\theta)}.
\eqno(2.28)
$$
It is easily seen that $a\theta=A$. On the other hand, In
accordance with {\it Key-formula } (2.12) we have
$$
\frac{c+a}{c-a}=\exp(2a\phi),~~
c=a\coth(a\phi),~~b=\frac{a}{\sinh(a\phi)}.   \eqno(2.29)
$$

\section{Pythagoras theorem and two dimensional  Fermi-like  oscillator}

 For the sake of convenience in this section let us use for derivatives short notations
$$
\frac{d}{d\phi}=d, ~~~\frac{d}{d\theta}=\partial.
$$
Then differentiating formula (2.22) and (2.23) we come to the
following system of differential equations
$$
d p_0=-p^2,~~~d p=-pp_0,~~~\partial p_0=-pp_0,~~~\partial
p=-p^2_0. \eqno(3.1)
$$
The operators $d$ and $\partial$ do not commute. Introduce two dimensional  vector of a state by
$$
\Phi(p_0,p)=\left( \begin{array}{c}
p_0\\
p
\end{array} \right). \eqno(3.2)
$$
Calculate actions of the operators $d^2-\partial^2$ and $d\partial-\partial d$ on this vector:
$$
(d^2-\partial^2)\Phi(p_0,p)=m^2~\Phi(p_0,p), \eqno(3.3)
$$
$$
(\partial d-d\partial)\Phi(p_0,p)=m^2~ \left( \begin{array}{cc}
0&1\\
1&0
\end{array} \right)\Phi(p_0,p). \eqno(3.4)
$$
Introduce operators
$$
a^-=d-\partial,~~a^+=d+\partial,
$$
with following commutation and anti-commutation rules
$$
\frac{1}{2}(a^-a^++a^+a^-)=m^2~
 \left( \begin{array}{cc}
1&0\\
0&1
\end{array} \right), \eqno(3.5)
$$
$$
a^-a^+-a^+a^-=2m^2 \left( \begin{array}{cc}
0&1\\
1&0
\end{array} \right). \eqno(3.6)
$$
It is seen, we deal with some kind of two dimensional Fermi-like oscillator with Hamilton operator
$$
H=\frac{1}{2}(a^-a^+-a^+a^-), \eqno(3.7)
$$
and with anti-commutation relation given by
$$
(a^-a^++a^+a^-)=2m^2, \eqno(3.8)
$$
acting on the  state $ \Phi_0(p_0,p)$. The Hamilton operator possesses with two eigenvalues
$$
H\Phi_n=E_n\Phi_n,~n=1,2, \eqno(3.9)
$$
where
$$
E_1=+m^2,~~E_2=-m^2,
$$
so that,
$$
a^+\Phi_1=\Phi_2,~~a^-\Phi_2=\Phi_1,
$$
and, due to anti-commutation relations (3.8),
$$
a^{\pm}H+Ha^{\pm}=0.
$$

\section{ Klein-Gordon equations  for energy-momentum of classical relativistic particle in the space of rapidity }

It is seen that equation (3.3) is nothing else than two
dimensional Klein-Gordon equation. Comparing this equation with
two dimensional Klein-Gordon equation written in terms of
space-time coordinates we come to conclusion that the parameter
$\phi$- is a time-like parameter, whereas the parameter $\theta$
is an analogue of a space coordinate. In order to pass to the
Klein -Gordon equation in four-dimensional Minkowski space with
signature $(+---)$ we shall extend the parameter $\theta$ till to
three dimensional vector. In this way we arrive to covariant
formulation of evolution equations.

The momentum is a spatial part of the four-vector energy-momentum
with components $p_k,k=1,2,3$. Now, instead of $\phi$ we will use
the letter $\rho_0$, and $\theta$ has to be replaced by spatial
part of four-vector of rapidity containing components
$\rho_1,\rho_2,\rho_3$.

In these variables the evolution equations have to be written in a Lorentz-covariant form. The evolution equations we
shall extend as follows. The single variable $p$ is replaced by the components of three-vector of momentum,
$p_k,k=1,2,3$.  The square $p^2$ means $p^2=-p^kp_k$. In this way we arrive to the following set of equations
$$
(a)~~~d^0~p_0=~-p^kp_k,~~~(b)~~d^0~p_k=p_kp^0,~~k=1,2,3.
\eqno(4.1)
$$
$$
(a)~~~\partial_k~p_0=p_kp_0,~~~(b)~~~~\partial^k~p_k=-p^2_0.
\eqno(4.2)
$$
Hereafter we use the following notations for derivatives
$$
\partial^k=\frac{\partial}{\partial\rho_k},~~~d^0=\frac{d}{d\rho_0},
$$
and  adopt, so-called, {\it a summation convention}, according to which any repeated index in one term, once up, once
down, implies summation over all its values.

 Remember, however, that there exist some functional dependence between $\rho_0$ and $\rho_k,k=1,2,3$, so that the
spatial variables are functions of the time-like parameter, i.e., $\rho_k=\rho_k(\rho_0),k=1,2,3.$ This means, the full
derivative with respect to $\rho_0$ is
$$
d^0p_0=-p^kp_k=~\frac{d\rho_k}{d\rho_0}\frac{\partial}{\partial
\rho_k} p_0. \eqno(4.3)
$$
On making use of equations (4.1)-(4.2), we get
$$
d^0p_0=p^2=-p^kp_k=~p_k\frac{d\rho_k}{d\rho_0}~p_0. \eqno(4.4)
$$
In order to provide this equality we have to take
$$
~p_k\frac{d\rho_k}{d\rho_0}=\frac{p^2}{p_0}. \eqno(4.5)
$$
Our purpose is to complete the evolution equations (4.1)-(4.2)
with an equation containing the following derivative
$$
~~\frac{\partial}{\partial \rho_n} p_k.
$$
 For that reason let us re-write equation (4.1b) as follows
$$
~\frac{d}{d\rho_0}~p_k=~\frac{d\rho_n}{d\rho_0}\frac{\partial}{\partial \rho_n}~ p_k=p_kp_0.
$$
In order to provide this equality we have to suppose that
$$
~\frac{\partial}{\partial \rho_n} p_k=p_kp^n\frac{p_0^2}{p^2}.
\eqno(4.6)
$$
One may easily check that formula (4.6) is in accordance with
(4.1) and (4.2).

{\bf Equations with second order derivatives.}

Firstly, calculate the second order derivatives of $p_0$ and $p$  with respect to time-like variable $\rho_0$. We have,
$$
\frac{d}{d\rho^0}\frac{d}{d\rho_0}p_0= -2p^kp_kp_0=2p^2p_0.
\eqno(4.7)
$$
Secondly, calculate action of the Laplace operator on $p_0$. Define the Laplace operator by
$$
\Delta=\frac{\partial}{\partial \rho_k}\frac{\partial}{\partial \rho^k}.
$$
By taking into account (4.2a)  we obtain
$$
\partial^k\partial_k~p_0=-p_0^2p_0+p^kp_kp_0=-p_0^3-p^2p_0. \eqno(4.8)
$$
Joining this equation with (4.7) we come to Klein-Gordon equation
for $p_0$:
$$
d^0d_0~p_0+\partial^k\partial_k~p_0=-m^2~p_0. \eqno(4.9)
$$

Now calculate action of operator $\Delta$ on $p_k$ by  using
formulae (4.6) and (4.2b).
$$
\partial^n\partial_n~p_k  =\partial^n~(~ p_kp_n\frac{p_0^2}{p^2}~)
$$
$$
=(\partial^n~(~p_kp_n))~\frac{p_0^2}{p^2}+p_kp_n\partial^n(~\frac{p_0^2}{p^2})
$$
$$
=\frac{p_0^2}{p^2}~(p^np_n)~\frac{p_0^2}{p^2}+p_k(\partial^np_n)~\frac{p_0^2}{p^2}~+p_kp_n\partial^n(~\frac{p_0^2}{p^2})
$$
$$
=-2\frac{p_0^4}{p^2}p_k~+p_kp_n\partial^n(~\frac{p_0^2}{p^2})
$$
$$
=-2\frac{p_0^4}{p^2}p_k~~-2p_kp^n\frac{p_np_0^2m^2}{p^4}
$$
$$
=-2\frac{p_0^4}{p^2}p_k+~~2p_k\frac{p_0^2m^2}{p^2}
$$
$$
=(~~-2\frac{p_0^4}{p^2}p_k+~2p_k\frac{p_0^4}{p^2}~~) =-2p_kp_0^2.
$$
Joining this result with
$$
\frac{d}{d\rho^0}\frac{d}{d\rho_0}~p_k=p_kp_0^2-p_kp^np_n=p_kp_0^2+p_kp^2,
$$
we come to analogue of Klein-Gordon equation for $p_k$:
$$
\Delta~p_k+d^0d_0~p_k~=-m^2~p_k.
$$

{\bf Comparison with Klein-Gordon  equation used in relativistic
quantum mechanics.}

From formula (4.5)
$$
~p_k\frac{d\rho^k}{d\rho_0}=\frac{p^2}{p_0},
$$
we may conclude that
$$
\frac{d\rho^k}{d\rho_0}=v^k+M^{kl}p_l,
$$
where
$$
v^k=\frac{dx^k}{dx^0}
$$
is the velocity with respect to coordinate time, $M^{kl}=-M^{lk}$ is an arbitrary anti-symmetric tensor.

In the relativistic quantum mechanics the Klein-Gordon equation  is obtained simply by using some conventional receipt
according to which components of four-momentum in the mass-shell equation are replaced by corresponding differential
operators as follows \cite{Ryder}
$$
p_k=-i\hbar\frac{\partial}{\partial x^k},~p_0=i\hbar\frac{\partial}{\partial x^0}.
$$
So, we  come to the following correspondence
$$
\hbar\rho^{\mu}\Rightarrow x^{\mu},~~\frac{\partial}{\partial \rho^{\mu}}=\hbar\frac{\partial}{\partial x^{\mu}}.
$$

\section{Concluding remarks}

1. We considered two ways of description an evolution constrained
 by Pythagoras formula. The first one is given by the hyperbolic angle, and the
 second one,  by the periodic angle. The both angles are proportional to
 the fixed side of the right angled triangle. In the case of
 relativistic mechanics, the hypotenuse is the energy, the fixed
 side is the mass and the moving side is the momentum.\\

2. The derivative of the periodic angle with respect to the
hyperbolic angle is equal to a ratio of the moving side to the
hypotenuse, in the case of relativistic mechanics, this ratio is
the velocity
$$
\frac{d\theta}{d\phi}=\frac{p}{p_0}=\frac{v}{c}.
$$

3. This relationship prompts us to conclude that the hyperbolic
 angle $\phi$ is the time-like parameter, whereas the periodic
 angle $\theta$ is the space-like parameter.\\

4. The evolution equations admit an extension to the case of three
(or more) dimensions, however, in this case we could not find an
explicit expression for the momenta.\\

5. Near the point where the fixed side of the triangle ( mass)
becomes infinitesimal and according to the Klein-Gordon equation
it is conjectured that this motion will display features of the
wave motion.

\end{document}